\documentclass[conf]{new-aiaa}
\usepackage[utf8]{inputenc}

\usepackage{graphicx}
\usepackage{amsmath}
\usepackage[version=4]{mhchem}
\usepackage{siunitx}
\usepackage{longtable,tabularx}
\title{Keldysh-Lattice Boltzmann approach to quantum nanofluidics}
\author{Sauro Succi \footnote{Senior Researcher, IIT Center for Life Nano-Neuroscience at La Sapienza, sauro.succi@gmail.com}}
\affil{IIT Center for Life Nano-Neuroscience at La Sapienza, \\
Viale Regina Elena 291, 00161, Roma, Italy}
\author{Andrea Montessori}
\affil{Roma Tre University, Department of Civil, Computer Science and Aeronautical Technologies Engineering, Rome, 00146, Italy}

\begin{document}

\maketitle

\begin{abstract}
We present a mathematical and computational framework to couple the
Keldysh non equilibrium quantum transport formalism with a nanoscale
lattice Boltzmann method for the computational design of quantum-engineered
nanofluidic devices.
\end{abstract}

\section{Introduction}

The general trend of modern science and engineering towards miniaturization
has placed a strong premium towards the study of fluid phenomena at nanometric scales.
This tendency draws from many technological sources, aeronautics and aerospace, biomedical , chemical-pharmaceutical and  
energy/environment being just a few prominent ones.  
Given the vast amount of energy lost on frictional contacts, 
low friction is a paramount goal for the optimal design of
most micro and nano-mechanical devices involved in these applications.

According to continuum mechanics, the pressure gradient to push a given mass flow
across a channel of diameter $D$ scales like $D^{-4}$, a relation that speaks clearly for 
the difficulty of pushing flows across miniaturized devices: a ten-fold decrease in radius demands
a ten-thousand fold increase in the required pressure.
The above scaling derives from the assumption that the fluid molecules in contact with 
solid walls do not exhibit any net motion, the so called {\it no-slip} condition, because
the remain trapped in local corrugations of the solid wall.
This assumption is no longer valid whenever the size of the channel becomes 
comparable with the molecular mean free path and, more generally, whenever the 
fluid-solid molecular interactions can no longer be described in terms of simple mechanical collisions.
Whatever the driving mechanism, the onset of a non-zero fluid velocity 
at the wall (slip flow), $u_s \ne 0$,  is a much-sought effect, as it turns 
the hydrodynamic $D^{-4}$ barrier into a much more  manageable $D^{-2}$ dependence.
Slip flow is typically quantified in terms of the so called slip length $L_s=u_s/(du/dy)_w$ where
$(du/dy)_w$ is the velocity gradient at the wall.
Under no-slip condition $L_s$ is of the order of the molecular mean free 
path, i,e. about $1 \;nm$ for water, while suitably treated (geometrically or chemically) walls 
can reach up to $L_s \sim 10 \div 100 \;nm$ ,  which is comparable to the size of the 
nano-device, leading to a very substantial decrease of the effective 
viscosity (roughly speaking a fa $D/L_s$).
Achieving large slip lengths involves the nano-engineering of fluid-wall interactions such 
as to prevent fluid molecules from being mechanical trapped by nano-corrugations. 
This is usually pursued by clever geometrical and/or chemical coatings, so as to
promote near-wall repulsion between fluid and solid molecules (hydrophobic coating). 

However, in the recent  years, it has been argued and experimentally shown that  
unanticipated quantum-electro-mechanical interfacial phenomena can lead to 
a major reduction of frictional losses \cite{KAV21,KAV22}. 

In this paper we wish to discuss prospects designing suitable computational 
solvers embedding ab-initio non-equilibrium quantum mechanical phenomena within
a mesocale computational harness, capable of exploring the effects
of the aforementioned quantum transport phenomena at scales of experimental relevance. 

We begin by discussing the basic physical scenario.

\section{The physical set up}

We consider a fluid of water molecules flowing in a nano-channel, 
say a carbon nanotube (CNT) confined by carbon walls, say graphite or graphene.  
The flow of liquid water is driven by an external pressure gradient and
this energy supply is dissipated to the walls. 
However, at variance with the classical picture, whereby such dissipation would result
from mechanical collisions of the water molecules with solid molecules at the wall,
new interfacial couplings need to be taken into account.
The first observation is that at these nanoscopic scales, the charge 
fluctuations of the water molecules at the liquid/solid interface couple with the 
electronic degrees of freedom in the solid, via screened Coulomb electrostatics. 
Given that De Broglie wavelength of electrons in metals is typically
of the order of $10\;nm$, such coupling is indeed pretty plausible.
The fluctuations of water molecules also couple to phonons excited by 
the direct impact of the water molecules on the solid wall. 
This generates a "phonon wind" in the solid which can in turn trigger
a corresponding "electron wind" in the solid via electron-phonon scattering.
Both species finally lose energy to the solid crystal, which acts as a ultimate
energy sink.

A detailed analysis based on the quantum non-equilibrium Keldysh formalism,
shows that the interaction between these three "species", hydrons 
(the charge fluctuations in the liquid), electrons in the solid and phonons 
also in the solid, leads to a rich variety of frictional exchanges between 
the flowing water and the electron/phonon "fluids" in the solid.
\begin{figure}
\centering
\includegraphics[scale=0.4]{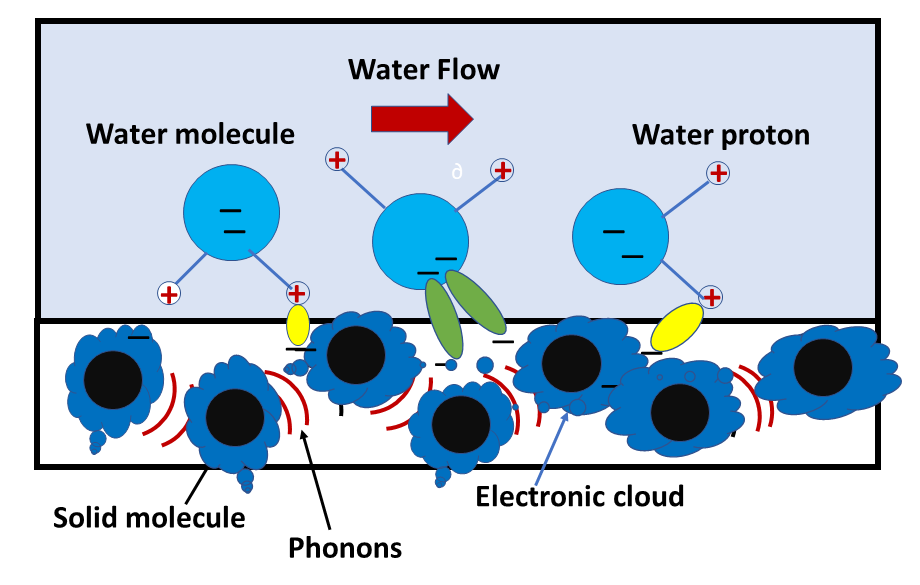}
\caption{The quantum friction scenario. The fluid part consists of water
molecules (light blue full circles) subject to charge fluctuations (hydrons).
The solid part consist of a crystal (full black circles), valence electrons and
lattice vibrations (phonons).  
The ovals represent coupling between charge fluctuations and electronic degrees
of freedom in the solid. Depending on the sign and orientation of the
coupling, the hydrons in the liquid maybe dragged or accelerated, the latter
instance resulting in drag reduction on the liquid water.
}
\end{figure}
Interestingly such exchanges can lead to a number of fascinating effects, such as
a sizeable reduction of the effective friction experienced by the water liquid, the 
so called "negative quantum friction" effect, and also "electronic current drive", namely
a net flow of electrons in the solid triggered by quantum interfacial effects.

Remarkably, the Keldysh analysis provides an analytical expression of these frictional effects
in terms of classical-looking Langevin terms of the form $\gamma_{kl}(u_k-u_l)$
where $k,l=1,3$ run over the three species in point, hydrons, electrons and phonons.

This motivates the formulation of a mesoscale-continuum model capable of investigating
the consequences of the Keldysh analysis on the scale of the experimental device.  
\begin{figure}
\centering
\includegraphics[scale=0.30]{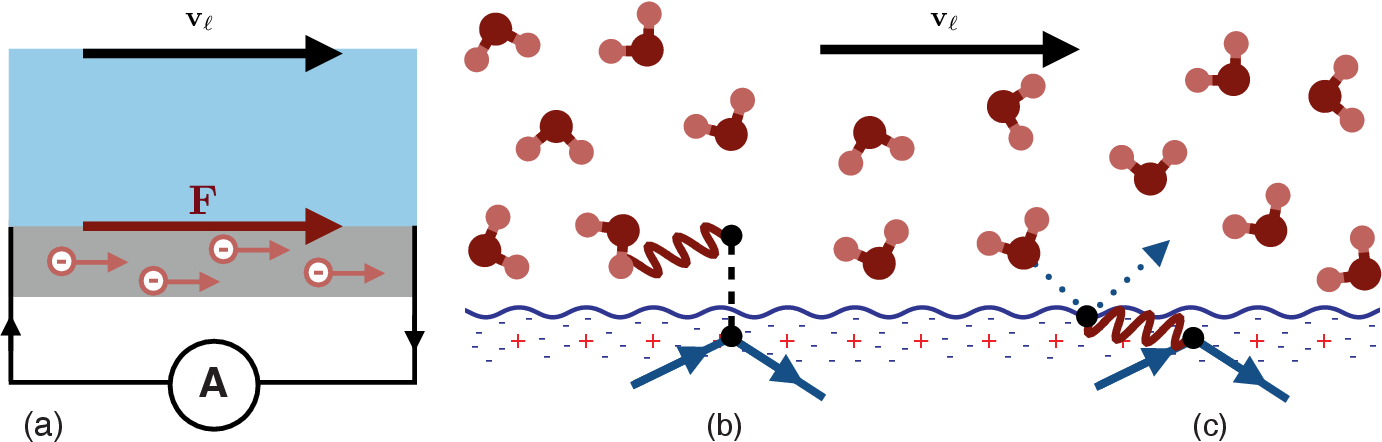}
\caption{Electron current-drive scenario.
(a) A nanofluidic device whereby liquid water drives an electronic current
in the solid, through two related but distinct mechanisms:
(b) Hydrons in liquid water impart momentum to the electrons in the solid through 
Coulomb interactions across the L/S solid interface. 
(c) Electrons in the solid are driven by phonons excited by water molecule 
collisions with the solid molecules.
From  Ref [3].
}
\end{figure}
\section{The HEP multi-fluid model}

In this section we portray a prospective fluctuating-hydrodynamic continuum model
of the physical set-up discussed in the previous section.
The idea is to import first-principle information from the Keldysh analysis at the
level of friction and fluctuating forces, to be inserted in the 
multi-fluid continuum model. 
\begin{figure}
\centering
\includegraphics[scale=0.35]{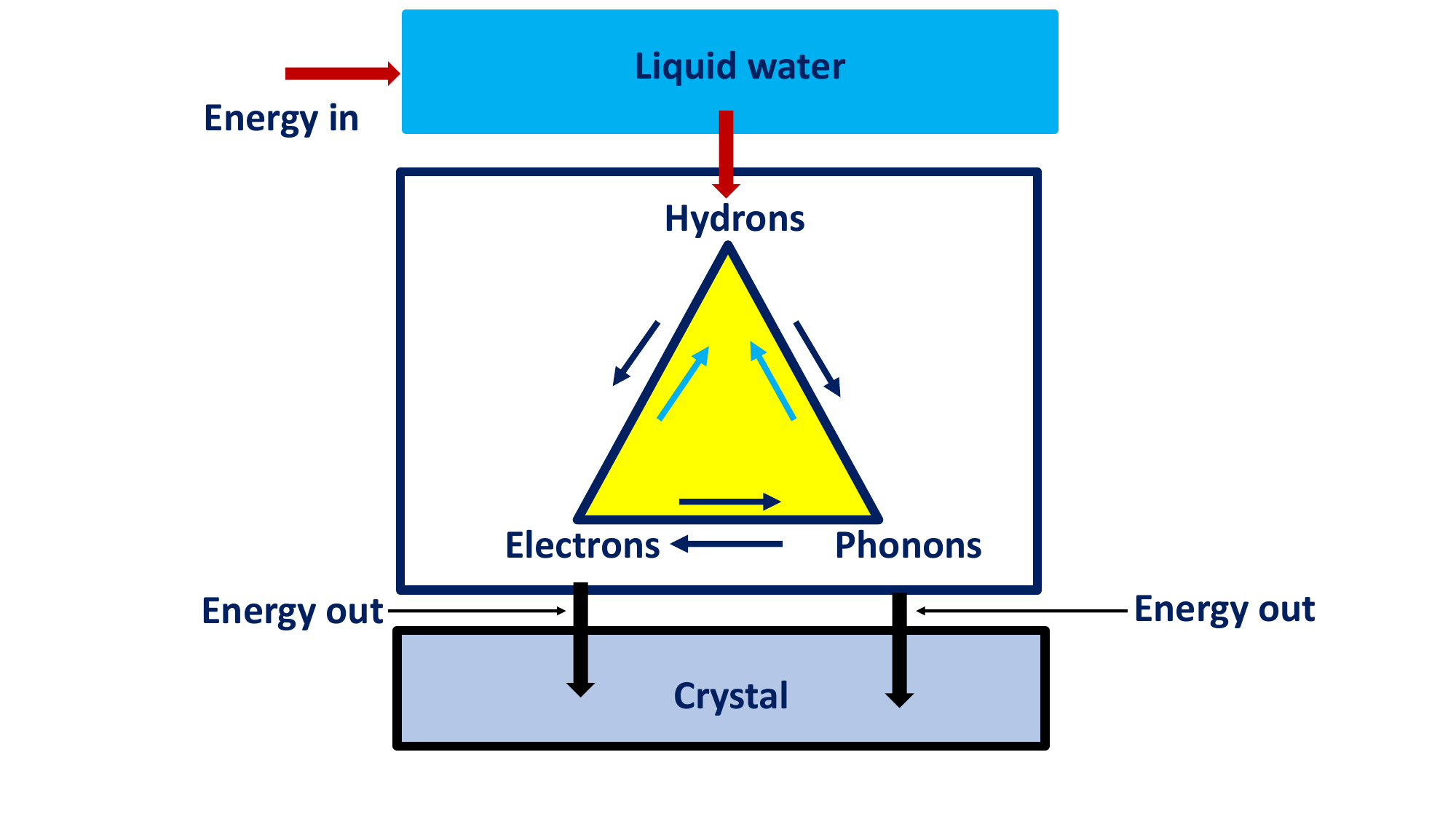}
\caption{The energy flow-chart of the HEP multi-fluid model.
The water flow (top) is driven by a pressure gradient and dissipates
energy at the liquid/solid interface.
Dissipation takes places through mutual momentum exchanges between the three
"fluid" species, hydrons (H), electrons (E) and phonons (P).
These interactions are governed by effective friction coefficients $\gamma_{kl}$,
$h,l=h,e,p$ which can go both ways depending on the physico-geometrical
conditions, primarily on the electron density.
The standard energy pathways is downward, hydrons attached to the water molecules
receive energy from the pressure gradient and they pass it on to the electrons
and phonons in the solid, thus generating a corresponding "phononic" and "electronic"
wind.Such winds are stabilized by frictional losses on the crystal atoms, which act
as an effective energy sink.  The internal pathways are however pretty rich.
In particular, when "inverse" couplings (upward flows back to the hydrons) are
activated, energy flows back into the hydrons, which is the hallmark 
of negative quantum friction. Depending on the geometrical and physical parameters
such reduction can be pretty substantial, up to fifty percent or more. 
}
\end{figure}

\subsection{The water fluid}
 
The water fluid obeys the standard continuity plus Stokes equations.
\begin{equation}
\partial_t n_w + \partial_a (n_w u_a) =0
\end{equation} 
\begin{equation}
\partial_t (n_w u_a)  + \partial_b (n_w u_a u_b + p \delta_{ab} + \sigma_{ab}) = F_a + F^h_a  
\end{equation} 
where $F_a$ are bulk forces while $F^h_a$ is the set of applied forces
due to interfacial inter actions.
Note that, despite the very low Reynolds numbers, we have retained the inertial
term, since such term is responsible for important nanoscale hydrodynamic correlations.

\subsection{Hydrons}

The hydrons move along with the water fluid, possibly with diffusive 
effects, hence they obey the continuity-diffusion equation
\begin{equation}
\partial_t n_h + \partial_a (n_h u_a) = D \partial^2 n_h
\end{equation} 
where $n_h$ is the hydron density and $u_a$ is the hydron velocity, which coincides
with the velocity of the water fluid.
The hydrons influence the motion of the carrier fluid, in which they are immersed,  via the body force term $F_a^h$ which embeds four separate contributions:
1) electrostatic contribution due to the interaction of the 
hydrons with the electrons in the solid, 2) Friction effects due to the
interaction with the phonons and electrons in the lattice,
3) Classical fluctuating forces, 4) Quantum fluctuating forces.
The electrostatic term it takes the form
\begin{equation}
F^{h,es} = -\nabla \; (z_h z_e n_h(x) \int V(x,y) n_e (y) dy
\end{equation}
where $z_h$ is the hydron charge (both positive or negative depending on whether
the charge fluctuations is due mostly to water protons or electrons) and $z_e=-1$
is the electronic charge. In the above $V(x,y)$ is the Debye-screened Coulomb potential.
Next, there is the aforementioned frictional drag (vector indices relaxed for simplicity)
\begin{equation}
F^{h,drag} = - m_h \gamma_{he}(u_h-u_e)- m_h \gamma_{hp}(u_h-u_p)  
\end{equation}
where $u_e$ and $u_p$ denote the net velocity of the
 electron and phonon "fluids" in the solid and $,m_h$ is the hydron mass.

Since we are dealing with nanoscale transport, friction must be complemented with
fluctuating forces, both thermal and quantum.

According to fluctuating-hydrodynamics, the thermal forces can be expressed as 
the divergence of the fluctuating momentum flux tensor, 
as defined by the Fluctuation-Dissipation Theorem:
\begin{equation}
\label{CFDT}
P_{ab}^{cf} = \langle \tilde F_a(x_1,t_1) \tilde F_b(x_2,t_2) \rangle
= \gamma_0 m k_BT \lambda^3 \delta(x_1-x_2) \delta(t_1-t_2)  \delta_{ab}
\end{equation}
where $\gamma_0$ is the classical collision frequency 
and $\lambda=v_{th}/\gamma_0$ is the mean free path. 

Quantum fluctuations may be treated similarly, but require additional care
since, depending on the energy of the quantum excitations, they 
may involve non-local effects in space and time, depending on their dispersion relation.  
\begin{equation}
\label{QFDT}
P_{ab}^{qf} = \langle \tilde F_a(x_1,t_1) \tilde F_b(x_2,t_2) \rangle
= \gamma_0 m k_BT \lambda_q^3 \Gamma_{ab}(x,t;(\frac{x_2-x_1}{\lambda_q},\frac{t_2-t_1}{\tau_q})  
\end{equation}
In the above, we have defined $x=(x_1+x_2)/2$, $t=(t_1+t_2)/2$ and
\begin{equation}
\tau_q = \frac{\hbar}{k_BT}
\end{equation}
as the typical time scale of thermal quantum fluctuations and $\lambda_q = \hbar/mv_{th}$
as the typical spatial scale (De Broglie length) while
$\Gamma_{ab}$ is the corresponding memory kernel

In the limit $\omega \tau_q \ll 1$ (soft modes), thermal fluctuations 
prevail and quantum excitations can be treated classically.
However, for "hard" quantum fluctuations with $\omega \tau_q > 1$, this is no
longer the case. 
Of course, a realistic estimate of quantum memory effects depends on the specific of the 
dispersion relation. For instance, hard excitations have usually shorter lifetime
than soft ones, hence quantum memory effects are expected mostly from the soft modes. 
For those modes, the momentum flux tensor may become non-local and the effective force
associated with quantum fluctuations can no longer be expressed as the divergence
of a one-point tensor.  


\subsection{Electron wind in the solid}

The treatment is basically the same as for the hydrons, a continuity equation 
plus a momentum equation supplemented with, i) electrostatic electron-electron and electron-hydron
screened Coulomb interactions, ii) friction with phonons and iii) the crystal atom.
Fluctuating forces may not needed because the electron orbitals are anchored to
the lattice atoms.
In equations:
\begin{equation}
F^{es,ee} = -\nabla (n_e(x) \int V_{ee}(x,y) n_e(y) dy 
\end{equation}
\begin{equation}
F^{es,eh} = -\nabla (n_e(x) \int V_{eh}(x,y) n_h(y) dy
\end{equation}
\begin{equation}
F^{fr,eh} = -m_e \gamma_{eh} (u_e-u_h) 
\end{equation}
\begin{equation}
F^{fr,ep} = -m_e \gamma_{ep} (u_e-u_p) 
\end{equation}
\begin{equation}
F^{fr,eC} = -m_e \gamma_{eC} u_e 
\end{equation}

It should be noted that interfacially-driven electrons in the solid move 
at about half the sound speed of phonons, roughly $1000$ m/s, hence they 
don't require a relativistic treatment as do electronic quasi-particles
in graphene \cite{Geim,Mendoza}. This reflects the fact that the electron wind
generated by quantum interfacial effects is very different from electron
conduction in graphene (where conduction electrons propagate
at about $c/100$).
Also to be noted that the water flow is several orders of magnitude 
slower, typically $u_w \sim 1$ $\mu s/m$, reflecting the vast mass separation
between the three species in point.

\subsection{Phonon wind in the solid}

Likewise, the "phonon fluid" in the solid obeys continuity and momentum
equations, the latter supplemented with frictional effects 
due to phonon scattering with electrons and the crystal atoms.
Phonons are linear waves associated with the lattice vibration, their
density (occupation number in quantum terms) is proportional to the amplitude of 
such oscillations and  obey a linear-damped wave equation:
\begin{equation}
\partial_{tt} n_p -c_s^2 \partial_{xx} n_p = -\gamma_p \partial_t n_p
\end{equation}
where the frictional force is the sum of phonon-electron 
and phonon-crystal scattering, namely:
\begin{equation}
F^{fr,pe} = -m_p \gamma_{pe} (u_p-u_e) 
\end{equation}
\begin{equation}
F^{fr,pC} = -m_p \gamma_{pC} u_p 
\end{equation}
$m_p= \hbar k/u_p$ being the effective phonon mass.  
As a result, $\gamma_p = \gamma_{pC} + \gamma_{pe}(1-u_e/u_p)$.

\section{The Keldysh Lattice Boltzmann (KLB) implementation}

The above fluctuating multi-fluid continuum model could be discretized with
grid methods for fluids. 
In the following we discuss a Lattice Boltzmann implementation \cite{HSB,succi2018} on account of 
the several advantages experienced for the case of microfluidics, and particularly
to represent fluid/solid interactions in a way which can seamlessly incorporate 
microscopic information without taxing computational efficiency.  
In addition, by using high-order lattices \cite{HOL}, LB can also accommodate classical
non-equilibrium effects beyond the realm of continuum mechanics, a statement that
holds true also for quantum non-equilibrium transport phenomena.
In this respect, the use of novel thread-safe, high-order LB models with reduced memory footprint \cite{GPU} will be instrumental to efficiently capture the complex non-equilibrium physics of the phenomena at hand without compromising the computational efficiency and scalability of LB models.

Indeed, with the set of fluctuating-hydrodynamic equations discussed in the previous section,
the LB implementation can proceed according to well established procedures.

There is a set of discrete distributions for each of the three species involved,
say $\lbrace h_i,e_i,p_i \rbrace$ and the corresponding interactions can 
be encoded according to the standard LB procedures to incorporate external/internal sources.
The local equilibria are second order expansions of the flow field of
the Bose-Einstein distribution for the bosons (hydrons and phonons) and 
Fermi-Dirac distribution for the electrons.
The implementation of thermal fluctuations is based on the standard fluctuating
LB formalism, while quantum fluctuations may require additional care.

Indeed, as it is well known, the classical fluctuating lattice Boltzmann equation
incorporates the fluctuating tensor in local form, through the following source term:
\begin{equation}
S_i(x,t) \propto P_{ab}(x,t) w_i (c_{ia}c_{ib}-c_s^2 \delta_{ab})
\end{equation}
where $P_{ab}$ fulfills the classical FDT relation \cite{Laddun}.

In the presence of a two-point quantum-fluctuating tensor 
$P_{ab}^{qf}(x,x+r;t,t+s)$, the corresponding non-local terms 
must be implemented.
Since this is very expensive, effective one-point closure could be attempted.
A possible path towards this end is to take weighted averages over the 
the running variables $r$ and $s$.
One could then define a one-point quantum fluctuating tensor as follows:
\begin{equation} 
\label{HEU}
P_{ab}^{qf}(x,t)= \sum_r \sum_s W_x(r) W_t(s) P_{ab}^{qf}(x,r;t,s)
\end{equation} 
with the weight functions suitably patterned after the two-point 
correlator $\Gamma_{ab}$, again an output of the Keldysh analysis.

\subsection{Estimating quantum memory effects}

It is useful to inspect some typical orders of magnitude of 
the main variables at play. 
The typical quantum thermal frequency is $\omega_{q}= k_bT/\hbar \sim 40$ ThZ,
meaning that the thermal quantum collision time is of the order of $0.1 \; ps$ 
(we remind that $\hbar \sim 6.5 10^{16}$ $ eV \cdot sec$ and $k_BT \sim 1/40$ $eV$).
With a LB timestep of, say, $1 ps$, no memory effects need be taken into account
since $\Delta t \gg \tau_{qt}$. Hence, the continuum Dirac delta can be safely
replaced by its lattice version $\frac{1}{\Delta t} (H(t)-H(t+\Delta t))$,
where $H$ denotes the Heaviside step-function.
The corresponding quantum thermal energy is about 25 $meV$, which is
comparable with the phonon and electron energies in play, but nonetheless below
the energy of the hard modes, which can reach up to a few hundreds $meV$,
namely $\hbar \omega/k_BT \sim 10$.    
While the details depend on the specific dispersion relation of the quantum excitations,
there are reasons to believe that memory effects should be taken into account through
informed heuristic relations.

\subsection{Boundary conditions}

Since there is flow on both liquid and solid sides of the nanodevice, it seems
natural to implement free-slip boundary conditions at the L/S interface
for all three species. 
Denoting by $(x_b,y_b)$ a generic site on the L/S boundary, for each site
on the boundary domain we have: 
\begin{eqnarray}
h_2(x_b,y_b) = h_4(x_b,y_b+1)\\
h_5(x_b,y_b) = h_7(x_b+1,y_b+1)\\
h_7(x_b,y_b) = h_5(x_b-1,y_b+1)
\end{eqnarray} 
where the discrete speed are numbered according to the standard D2Q9 notation
(1=right,2=up,3=left,4=down,5=right-up,6=left-up,t=left-down,8=right-down).

Likewise, for the electrons and the phonons ($f=e,p$), we have:
\begin{eqnarray}
f_4(x_b,y_b) = f_2(x_b,y_b-1)\\
f_7(x_b,y_b) = f_5(x_b-1,y_b-1)\\
f_8(x_b,y_b) = f_6(x_b+1,y_b-1)
\end{eqnarray} 

Once could also possibly implement electron and phonon scattering from the solid atoms
via the standard bounce-back no slip boundary conditions, but it is not obvious that this
would be an adequate description of such interactions.

At the external upper and lower solid boundaries one can implement no-slip flow, i.e. the
bounce back rule, as long as the molecular mean free path remains well below the
crossflow direction of the nano-channel.
Inlet and outlet boundaries can be handled either by periodic boundary conditions or via Neumann-like boundary conditions as developed in \cite{GPU2}, the latter possessing the advantage to be extended to higher-order lattices without any additional algorithmic effort.

\begin{figure}
\includegraphics[scale=0.4]{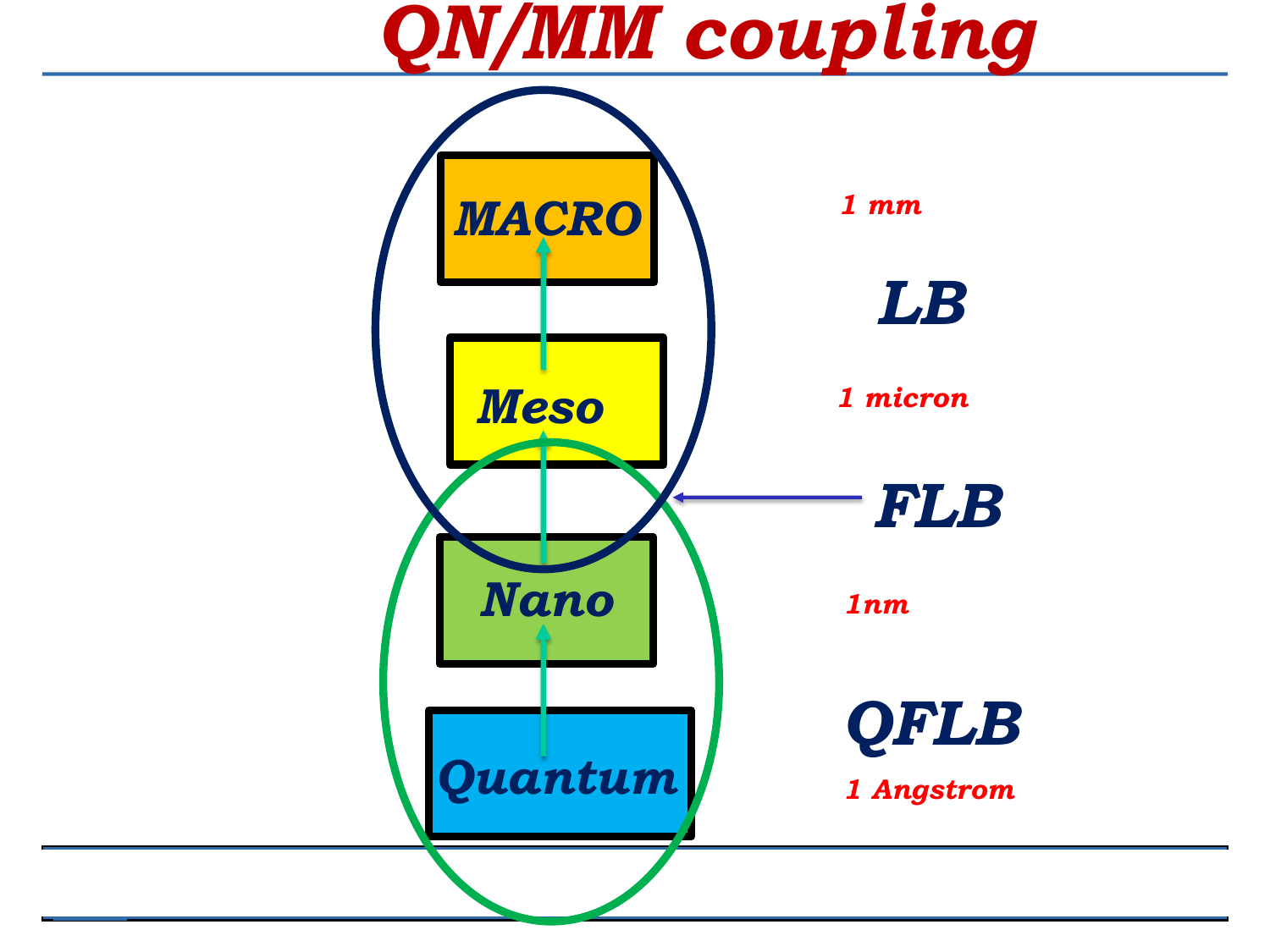}
\centering
\caption{Schematics of the KLB approach within the multiscale ladder.
Lattice Boltzmann was born to address macroscopic fluids, conventionally above
$1$ mm in the figure and subsequently it was extended to meso-microscale fluids.
With the inclusion of fluctuating forces it was brought to nanoscales (FLB).
Keldysh Lattice Boltzmann (KLB) goes one level further down, into the quantum
non equilibrium realm.
This is less surprising than it seems if one considers two "simple" and yet
non trivial pieces of evidence. 
First, at least for simple fluids, the continuum description may hold down to a few nanometers.
second, the de Broglie wavelength of thermal electrons in metals is about $10\;nm$.
Hence the quantum and continuum descriptions can definitely shake hands, and the
Keldysh Lattice Boltzmann approach just aims at exploiting this overlap.   
}
\end{figure}

\section{Prospective applications}

As mentioned earlier on, the main scope of KLB is to incorporate genuine non-equilibrium
quantum information within a  computational harness capable of exploring its effects
on a scale of experimental interest.
This is of decided interest, because the Keldysh analysis is performed under
a number of theoretical restrictions, for instance translational invariance, which
are clearly broken when realistic nanodevice geometries are taken in consideration.
This is precisely where LB is supposed to bring its added value.

The KLB would allow the quantitative analysis of a number of very interesting
applications, centred on the design of quantum-controlled ultra-low friction nanodevices.
In particular, one could probe negative quantum friction scenarios at scales
of direct experimental relevance. Below, we sketch three prospective possibilities.

\subsection{Water permeability in carbon nanotubes}

It has been found that the permeability
of water scales like the inverse radius of the CNT, a finding which escapes 
molecular dynamics simulations, indicating that the actual electronic configuration, and
not just crystallographic symmetries, dictates nanoscale water transport.
Indeed, no such effect is observed with graphite CNT, which share the 
same crystallographic symmetries of graphene. 
Likewise, it would of utmost importance to study the effects of nano-corrugations or wrinkles
of the graphene sheets on water transport, along the same lines successfully explored
by LB simulations of strongly confined microfluids.

The KLB tool could then be applied to the study of water transport in nanotubes, to benchmark
against existing treatments and extend the analysis to larger sizes in space and time. 
The goal is to parametrize quantum friction effects as a function of the design 
parameters, diameter and length of the nanotube, wall corrugations, 
and charge carrier density. 

\begin{figure}
\centering
\includegraphics[scale=1.4]{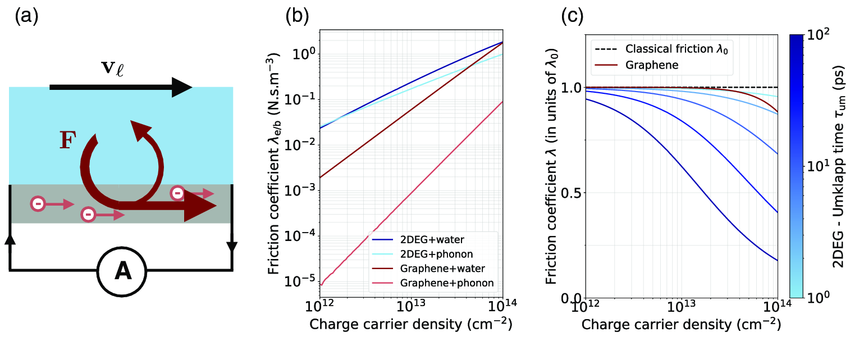}
\caption{Negative quantum friction resulting from momentum return from the solid to
the liquid via phononic excitations at the liquid/solid interface (from ref. [2].)
}
\end{figure} 
\subsection{Droplet-driven graphene currents} 

Recent experiments have shown that ionic-liquid droplet rolling over graphene sheets
can drive substantial electric currents due to the non-equilibrium coupling between droplet
charge fluctuations and graphene electrons \cite{KAV23a}.
The KLB tool could be used to simulate various experimental setups, 
i.e. varying droplet size and speed and the surface topography to identify the lowest
friction scenarios. 
KLB could enable multi-droplet design, by exploring the effects 
induced by nonlinear coupling between the electrons in the wake of the 
first droplet  with the charge fluctuations of the second one.
\subsection{Droplet-driven currents in twisted graphene} 

Electron super-conductivity in magic-angle twisted graphene is one of the most stunning
discoveries in modern material science \cite{MAGIC}. 
It is a genuinely electronic effect which is usually 
tackled by sophisticated electronic structure simulations.
However, whenever electrons couple to electro-nano-hydrodynamics, it is legitimate to
speculate that new electro-hydro driven electronic effects could arise and 
be observed via KLB simulations.
For instance, it would be interesting to investigate whether 
non-symmetric flow configurations induced by off-centered
charged droplet motions could enhance magic-angle-like effects
or perhaps suppress them.
This is just one more example of applications that might be pursued via KLB simulations.

\section{Computational feasibility}

We conclude with some assessment on the computational feasibility of the KLB scenario.

The natural tool of the trade for computational nanofluidics is Molecular Dynamics 
\cite{Petros1,Petros2}.
Despite the tremendous progress since its inception, MD still falls short of reaching up to the
scale of experimental interest. For instance, a (100, 1000) carbon nanotube ($100\;nm$ diameter and
$1000\;nm$ length) contains about $300$ millions water molecules. With order $10\;Kflops$ per molecule
and timestep, this makes about 3 trillions $flops/step$: a top-of-the line 30 (effective) $Petaflops/s$ 
supercomputer would complete a single timestep in about $100$ microseconds. 
Ticking at  1 fs ($10^{-15}$ s) timestep, this supercomputer would 
evolve the full system over a time span of about $1$ microseconds/day.
Various sources of inefficiency bring this estimate down to about $10 \div 100 ns/day$, which falls
short of meeting engineering design criteria by two orders of magnitude. 
One way out is the resort to hybrid hydro-molecular dynamics (HMD), whereby MD would be
confined to the near-wall region, leaving the bulk to a hydrodynamic solver.
This is however still demanding and subject to a series of physical on computational
open questions \cite{Delgado}.
More importantly, it has been shown that slip flow in CNT is largely underestimated by MD,
indicating that standard force-field procedures fall short of describing the actual physics
of the water-graphene interactions, pointing to a crucial role of electronic 
degrees of freedom in the solid. Interestingly, ab-initio MD, besides being even more
unpractical computing-wise, would also fall short of capturing the basics of quantum friction because 
the electrons in the solid are non-adiabatically coupled to charge fluctuations in the liquid.  

With a lattice spacing $\Delta x =1 nm$, and a time-step $\Delta t=1ps$,  a nanotube of 
diameter $D=100\;nm$  and length $L=10D$  takes $100 \times 100 \times 1000 =10^7$ 
grid points, so that a device made of ten such nanotubes side by side would take 
$100$ million points, a modest size for today's LB standards. 
Extreme LB simulations (hundred billion grid points) can reach thousand times higher, permitting to
either increase the resolution or upscale the device by a factor ten along each direction.
At a processing speed of $100\; GLUPS$, $100\; billion$ lattice sites are 
updated every second,  hence one million steps (1 microsecond) take roughly 1Ms, about
two weeks wall clock time ($1000 \;ns/day$, versus $10 \;ns/day$ of the current leading edge MD simulations).  
So much for extreme simulations. 
Standard KLB could operate with one billion grid points, bringing
the wall clock time down to a few hours, which is totally compatible 
with engineering design requirements \cite{GPU}. 
To be noted that KLB operates basically at the same spatial resolution of MD, namely
$0.1 \div 1 \;nm$ , but on a much coarser time-scale.
The key point is that, while MD needs timesteps much shorter ($\Delta t/\tau_c \sim 1/100$) 
than the typical collision time in order to conserve energy, thanks to the exact streaming
and built in conservative collisions, LB affords stable operation with 
timesteps $\Delta t/\tau_c \sim 1$.

This shows that KLB is computationally feasible and may offer a very appealing tool
tool to inspire and guide a new generation of experiments suggesting 
new strategies to design low-friction nanofluidic devices at the quantum level.

\section{Summary}
 
Summarizing, we have sketched a prospective scenario to embed ab-initio non-equilibrium
quantum transport effects within a fluctuating continuum hydrodynamic 
model, to be implemented within a Lattice Boltzmann framework.

The task is intense but seemingly feasible, both conceptually and computationally.
As a potential major extension as compared to existing LB schemes we have pointed out the
prospect of implementing a non-local version of the fluctuation dissipation theorem
for long-memory quantum correlations, both in time and space.
The actual manifestation of such non-local effect is very likely to depend on the 
specific application in point, but in case the need arise, a prospective 
one-point effective formulation has been briefly discussed.
Needless to say, these are simply feasibility considerations, the real test
lies within actual simulations, hopefully to be tackled in the near future. 

\section{Acknowledgments} 
The author is grateful to L. Bocquet for introducing him to the beautiful subject
of quantum nanofluidics and for many invaluable insights.  
This work was triggered by the Symposium "La nanofluidique à la croisée des chemins",
College de France, Paris, May 2023.

\section{Appendix: The Keldysh analysis}

The Keldysh formalism is based on the dynamic equation for the 
Green function associated with the particle generation and destruction operators
$\hat \Psi$ and $\hat{\Psi^+}$ 
\begin{equation}
G_K(x_1,x_2;t_1,t_2) = < [ \hat \Psi(x_1,t_1) \hat \Psi(x_2,t_2) ]_{-} >
\end{equation}
where subscript $-$ denotes the anticommutator.
By setting $X=(x_1+x_2)/2$, $x = (x_1-x_2)/2$, $T=(t_1+t_2)/2$ and $t= (t_2-t_1)$
and taking the Fourier-transform, we obtain the Keldysh Green function in eight-dimensional
phase-spacetime $(X,P,E,T)$
\begin{equation}
G_K(X,T;P,E) = \int e^{-i(Px-Et)/\hbar} G(x,X;t,T) dx dt 
\end{equation}
Clearly, for the homogeneous case, the dependence on $X$ and $T$ drops out, but
since we shall be dealing with quantum non-equilibrium  transport phenomena, such an assumption
is not justified. 
The equation for $G_K$ can be derived from first principles, i..e for the underlying Schrodinger
equation in second quantized form. Understandably, this is overly complicated, and the fact
of living in eight-dimension  surely does not help.
Nevertheless, in a series of inspiring papers, the authors managed to derive
a  number of important insights based on analytical manipulations 
of the Keldysh Green function under the hypothesis of homogeneity..
The main result is that the three species experience a frictional drag linearly proportional
to their relative net flow velocity, just like in a classical Langevin framework.
The quantum non-equilibrium features remain hidden in the values of the friction coefficients, i.e.
\begin{equation}
F_{kl} = - F_{lk} = \gamma_{k,l} (u_k-u_l)
\end{equation} 
where $k,l=1,3$ run over the three "species" $h$ (hydrons), $e$ (electrons) and $p$ (phonons). 


\begin{thebibliography}{99}

\bibitem{KAV21} Kavokine, N., Netz, R. R., \& Bocquet, L. (2021). Fluids at the nanoscale: From continuum to subcontinuum transport. Annual Review of Fluid Mechanics, 53, 377-410.

\bibitem{KAV22} Kavokine, N., Bocquet, M. L., \& Bocquet, L. (2022). Fluctuation-induced quantum friction in nanoscale water flows. Nature, 602(7895), 84-90.

\bibitem{HSB} Higuera, F. J., Succi, S., \& Benzi, R. (1989). Lattice gas dynamics with enhanced collisions. Europhysics letters, 9(4), 345.

\bibitem{succi2018} Succi, S. (2018). The lattice Boltzmann equation: for complex states of flowing matter. Oxford university press.

\bibitem{KAV23a} Coquinot, B., Bocquet, L., \& Kavokine, N. (2023). Quantum feedback at the solid-liquid interface: Flow-induced electronic current and its negative contribution to friction. Physical Review X, 13(1), 011019.

\bibitem{KAV23b} Lizée, M., Marcotte, A., Coquinot, B., Kavokine, N., Sobnath, K., Barraud, C., ... \& Siria, A. (2023). Strong electronic winds blowing under liquid flows on carbon surfaces. Physical Review X, 13(1), 011020.

\bibitem{Delgado} Delgado-Buscalioni, R., \& Coveney, P. V. (2003). Continuum-particle hybrid coupling for mass, momentum, and energy transfers in unsteady fluid flow. Physical Review E, 67(4), 046704.

\bibitem{Petros1}  Werder, T. Walther, J. H., Jaffe R.L., Halicioglu, T. \& Komoutsakos, P., (2003). On the water-carbon interaction for use in molecular dynamics simulations of graphite and carbon nanotubes. The Journal of Physical Chemistry B, 107(6), 1345-1352.

\bibitem{Petros2} Papadopoulou, E., Kim, G. W., Koumoutsakos, P., \& Kim, G. (2023). Molecular dynamics analysis of water flow through a multiply connected carbon nanotube channel. Current Applied Physics, 45, 64-71.

\bibitem{Horbach} Horbach, J., \& Succi, S. (2006). Lattice Boltzmann versus molecular dynamics simulation of nanoscale hydrodynamic flows. Physical review letters, 96(22), 224503.

\bibitem{Fyta}  Fyta, M. G., Melchionna, S., Kaxiras, E., \& Succi, S. (2006). Multiscale coupling of molecular dynamics and hydrodynamics: application to DNA translocation through a nanopore. Multiscale Modeling \& Simulation, 5(4), 1156-1173.

\bibitem{Ladd1} Ladd, A. J. (1994). Numerical simulations of particulate suspensions via a discretized Boltzmann equation. Part 2. Numerical results. Journal of fluid mechanics, 271, 311-339.

\bibitem{Ladd2} Ladd, A. J. (1994). Numerical simulations of particulate suspensions via a discretized Boltzmann equation. Part 2. Numerical results. Journal of fluid mechanics, 271, 311-339.

\bibitem{Laddun} Dünweg, B., \& Ladd, A. J. (2009). Lattice Boltzmann simulations of soft matter systems. Advanced computer simulation approaches for soft matter sciences III, 89-166.

\bibitem{Amadei} Amadei, C. A., Montessori, A., Kadow, J. P., Succi, S., \& Vecitis, C. D. (2017). Role of oxygen functionalities in graphene oxide architectural laminate subnanometer spacing and water transport. Environmental Science \& Technology, 51(8), 4280-4288.

\bibitem{Geim} Geim, A. K., \& Novoselov, K. S. (2007). The rise of graphene. Nature materials, 6(3), 183-191.

\bibitem{Mendoza} Mendoza, M., Herrmann, H. J., \& Succi, S. (2013). Hydrodynamic model for conductivity in graphene. Scientific reports, 3(1), 1052.

\bibitem{MAGIC} Cao, Y., Fatemi, V., Fang, S., Watanabe, K., Taniguchi, T., Kaxiras, E., \& Jarillo-Herrero, P. (2018). Unconventional superconductivity in magic-angle graphene superlattices. Nature, 556(7699), 43-50.

\bibitem{HOL} Montessori, A., Prestininzi, P., La Rocca, M., \& Succi, S. (2015). Lattice Boltzmann approach for complex nonequilibrium flows. Physical Review E, 92(4), 043308.

\bibitem{GPU} Montessori, A., Lauricella, M., Tiribocchi, A., Durve, M., La Rocca, M., Amati, G., ... \& Succi, S. (2023). Thread-safe lattice Boltzmann for high-performance computing on GPUs. Journal of Computational Science, 74, 102165.

\bibitem{GPU2}  Montessori, A., La Rocca, M., Amati, G., Lauricella, M., Tiribocchi, A., \& Succi, S. (2024). High-order thread-safe lattice Boltzmann model for HPC turbulent flow simulations. \textit{arXiv preprint arXiv:2401.17074}. 

\end{thebibliography}

\end{document}